\documentclass{aa}  

\usepackage{graphicx}
\usepackage{txfonts}
\usepackage{natbib}
\usepackage{xspace}
\usepackage{color}
\usepackage{hyperref}

\usepackage{threeparttable}
\usepackage{supertabular}
\usepackage{multirow}
\usepackage{makecell}

\usepackage{amsmath}
\usepackage{subcaption} 
\usepackage{upgreek}
\usepackage{float}

\usepackage{tablefootnote}
\usepackage{tikz}
\usepackage{lscape}
\newcommand{\cref}{\ref}

\begin{document}

%\label{firstpage}

\title{NICER detection of a new candidate cyclotron line in the bursting X-ray pulsar GRO J1744-28}

\author{Qi Liu\inst{1,2}\fnmsep\thanks{qi.liu@hebtu.edu.cn}
\and Lingda Kong\inst{3}\fnmsep\thanks{lingda.kong@nankai.edu.cn}
\and Long Ji\inst{4}
\and Lorenzo Ducci\inst{2}
\and Xiaohang Dai\inst{5}
\and Andrea Santangelo\inst{2}
\and Ce Cai\inst{1}
\and Shuwang Cui\inst{1}
\and Mingyu Ge\inst{6}
\and Shu Zhang\inst{6}
\and Shuang-Nan Zhang\inst{6}
\and Lian Tao\inst{6}
\and Hua Feng\inst{6}
\and Wei Wang\inst{7}
\and Valery Suleimanov\inst{2}
\and Sergey Tsygankov\inst{8}
\and Juri Poutanen\inst{8}
\and Alexander Mushtukov\inst{9,10}
}
\institute{College of Physics, Hebei Normal University, Shijiazhuang 050024, China 
\and 
Institut für Astronomie und Astrophysik, Universität Tübingen, Sand 1, 72076 Tübingen, Germany 
\and 
School of Physics, Nankai University, Tianjin 300071, China
\and
School of Physics and Astronomy, Sun Yat-Sen University, Zhuhai 519082, China
\and
KeyLaboratory of Time and Frequency Primary Standards, Chinese Academy of Sciences, Xi’an 710600, China
\and
Key Laboratory of Particle Astrophysics, Institute of High Energy Physics, Chinese Academy of Sciences, Beijing 100049, China
\and 
Department of Astronomy, School of Physics and Technology, Wuhan University, Wuhan 430072, China
\and
Department of Physics and Astronomy, University of Turku, Turku FI-20014, Finland
\and
Mullard Space Science Laboratory, University College London, Holmbury St. Mary, Surrey RH5 6NT, UK
\and
Astrophysics, Department of Physics, University of Oxford, Oxford OX1 3RH, UK
}
\date{ }

\abstract
{We report the detection of cyclotron resonant scattering features (CRSFs) in the spectrum of the unique bursting pulsar GRO J1744-28, observed during its recent outburst in 2021 with the Neutron Star Interior Composition Explorer (NICER). Clear pulsations at a frequency of 2.141128 Hz as well as Type II X-ray bursts were observed. The pulse profile exhibits a single-peaked shape in all energy bands, with the pulse fraction showing a positive correlation with energy. We find that the persistent X-ray continuum of the accreting pulsar is well described by typical phenomenological models, and we confirm the presence of the cyclotron line at $\sim$5 keV as reported in previous studies. In addition, we detect a candidate absorption feature with a centroid energy of 2 keV. If confirmed, this feature could be interpreted as a CRSF, which would correspond to a magnetic field of $\sim$1.8 $\times 10^{11}$ G. Pulse-phase-resolved analysis also reveals this absorption line around the peak pulse phases. These NICER observations provide tentative evidence for the cyclotron line candidate, establishing GRO J1744-28 as a key laboratory for studying accretion physics in an intermediate-strength magnetic field.
}

\keywords
{stars: neutron - stars: magnetic field - pulsars: individual: GRO~J1744-28 - X-rays: binaries}

\maketitle

\section{Introduction}
\label{sec:introduction}

X-ray binary pulsars (XRBPs) consist of a highly magnetized neutron star and an optical companion. These systems are typically divided into high-mass (HMXBs) and low-mass X-ray binaries (LMXBs), based on whether the mass of the donor star is above $\sim$10~M$_{\odot}$ or below $\sim$2~M$_{\odot}$, respectively (\citealt{2011BASI...39..429P,2011Ap&SS.332....1R}; \citealt{2010AdSpR..45..949B}). High-mass X-ray binaries are further divided into supergiant and Be/X-ray binaries. They are young systems with strong magnetic fields ($10^{12}$--$10^{13}$~G), and accretion is typically wind-fed or occurs via a transient disk. In contrast, neutron stars in LMXBs are older and have weaker magnetic fields ($B \lesssim 10^9$~G). Accretion in LMXBs is usually disk-fed via Roche-lobe overflow. These systems often exhibit thermonuclear (type-I) X-ray bursts and can be spun up to millisecond periods, eventually evolving into radio millisecond pulsars (\citealt{1991PhR...203....1B}).

For accretion onto neutron stars, matter is funnelled to the magnetic poles by the magnetic field, where it is decelerated and forms hot spots or accretion columns, producing X-ray emissions. Cyclotron resonance scattering features (CRSFs; \citealt{1978ApJ...219L.105T, 2019A&A...622A..61S}) are observed as absorption-like features in the X-ray spectra of many pulsars. They arise from the quantization of electron energy levels in a strong magnetic field and provide a direct measurement of the magnetic field strength via the relation $E_{\mathrm{cyc}} = 11.57 B_{12} (1 + z)^{-1} \mathrm{keV}$, where $B_{12}$ is the magnetic field in units of $10^{12}$~G and $z$ is the gravitational redshift \citep{1992herm.book.....M}.

The bursting pulsar, GRO~J1744--28, was discovered in 1995 \citep{1995IAUC.6272....1F} and is located near the Galactic centre at a distance of $\sim$8~kpc \citep{1997ApJ...486.1013A,1999ApJ...517..436N}. It is unique in exhibiting both coherent X-ray pulsations at a spin frequency of 2.14~Hz \citep{1996Natur.379..799K} and prolific type-II X-ray bursts \citep{1996ApJ...462L..39L}—one of its peculiarities. These bursts typically last for several tens of seconds. GRO~J1744--28 has undergone five major X-ray outbursts since its discovery. The first occurred from December 1995 to May 1996 \citep{1995IAUC.6272....1F}. Approximately one year later, a second outburst began in December 1996 and lasted about four months \citep{2015MNRAS.452.2490D,2019MNRAS.482.1110J}. The third and brightest outburst was observed in January 2014, reaching a peak luminosity of $1.9 \times 10^{38}$~erg~s$^{-1}$, as monitored by \textit{NuSTAR}, \textit{Chandra}, \textit{XMM-Newton}, and \textit{INTEGRAL} \citep{2015MNRAS.449.4288D,2015ApJ...804...43Y}. The source entered a fourth, significantly fainter episode in February 2017, which was studied by \cite{2020A&A...643A.128K}. \cite{2020A&A...643A..62D} studied the X-ray emission of GRO J1744-28 in quiescence with \textit{XMM-Newton}. This paper focuses on the most recent, fifth outburst in 2021 June 04, which was observed by \textrm{NICER}. Swift/XRT observed the source on 2021 June 08 and estimated an unabsorbed flux (0.3-10 keV) of about 1.5 $\times 10^{-9} \rm erg/s/cm^{2}$.

Based on \textit{XMM-Newton} and \textit{INTEGRAL} observations of its 2014 outburst, a CRSF at 4.7~keV was reported, with indications of two possible harmonics \citep{2015MNRAS.449.4288D}. This is consistent with an earlier analysis of the 1997 outburst using \textit{BeppoSAX}, which suggested a CRSF at $\sim$4.5~keV \citep{2015MNRAS.452.2490D}. Both features imply an intermediate magnetic field strength of $B \sim 5 \times 10^{11}$~G. However, this field strength inferred from the $\sim$4.7~keV CRSF is significantly higher than estimates derived from other methods. These alternative estimates include accretion disk reflection modelling ($2-6 \times 10^{10}$~G), spin-up measurements ($\sim 9 \times 10^{10}$~G), and the propeller-effect flux threshold ($\sim 2 \times 10^{11}$~G) \citep{2014ApJ...796L...9D, 2015ApJ...804...43Y, 1997ApJ...482L.163C}.

\textrm{NICER} enables high-resolution, high-throughput spectroscopy in the soft X-ray band (0.2--12~keV) where such a CRSF is expected. Our goal is therefore to search for the cyclotron line with \textrm{NICER} and to perform a detailed phase-averaged and phase-resolved spectral analysis to probe the accretion geometry. In this work, we present a comprehensive spectral-timing analysis of GRO~J1744--28 using \textrm{NICER} observations from the pulsar's 2021 outburst. We report the detection of a CRSF at 5.0 keV. In addition, we find hints of a new CRSF at 2.0 keV, which is statistically significant, and briefly discuss the source's critical relevance for understanding accretion physics at intermediate magnetic field strengths. Section~\ref{sec:Data} describes the NICER observations and data reduction. The timing and spectral analysis are presented in Sect.~\ref{sec:results}. We discuss the implications of our results in Sect.~\ref{sec:discussion} and conclude with a summary of the paper.

\section{Data reduction} \label{sec:Data}

\begin{figure}
    \centering
    \includegraphics[width=.5\textwidth]{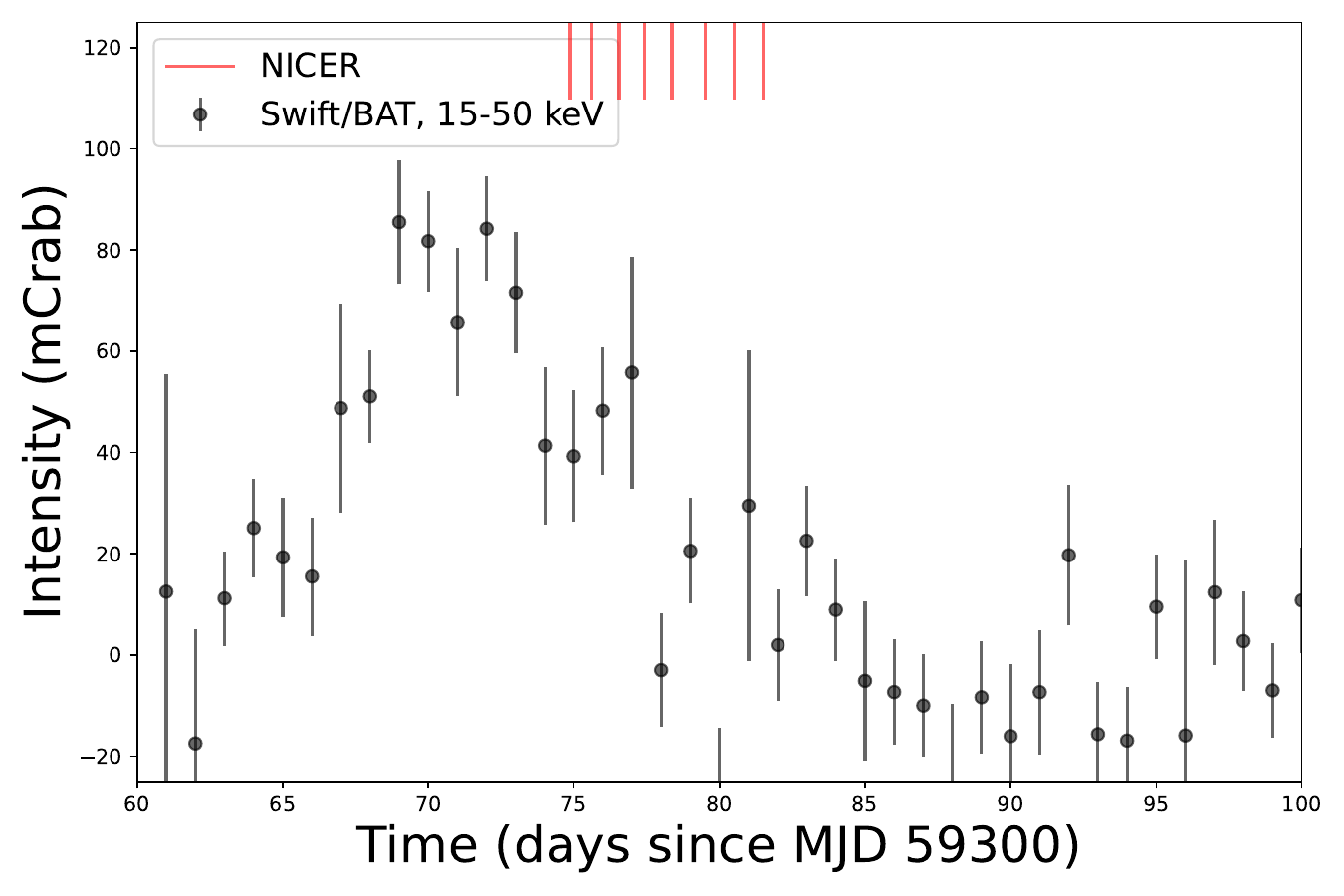}
    \caption{Light curve of the 2021 outburst via \textit{Swift}/BAT in the 15-50 keV energy range (black dots) from the NICER observations (vertical red lines).}
    \label{fig:counts}
\end{figure}

\begin{table}
    \centering
    %\footnotesize
    \caption{Observation IDs of NICER during the outburst
in 2021.}
    \label{tab:ObsIDs}
    \begin{tabular}{l|cccc}
    \hline \hline 
    \multirow{2}{*}{ObsID} & \multirow{2}{*}{\thead{Time Start \\ (UTC)}}& \multirow{2}{*}{MJD} & \multirow{2}{*}{\thead{Exposure time \\ (s)}}  \\ \\
    \hline 
4202210101 & 2021-06-09 & 59374.88 & 3125.0 \\
4202210102 & 2021-06-10  & 59375.63 & 2497.0 \\
4202210103 & 2021-06-11 & 59376.56 & 5475.0 \\
4202210104 & 2021-06-12 &  59377.44 & 3306.0 \\
4202210105 & 2021-06-12 &  59378.37 & 4347.0 \\
4202210106 & 2021-06-14 &  59379.53 & 4152.0 \\
4202210107 & 2021-06-15 &  59380.50 & 2893.0 \\
4202210108 & 2021-06-16 & 59381.50 & 4137.0 \\

    \hline
    \end{tabular} 
\end{table}

\begin{figure}
    \centering
    \includegraphics[angle=0,width=.5\textwidth]{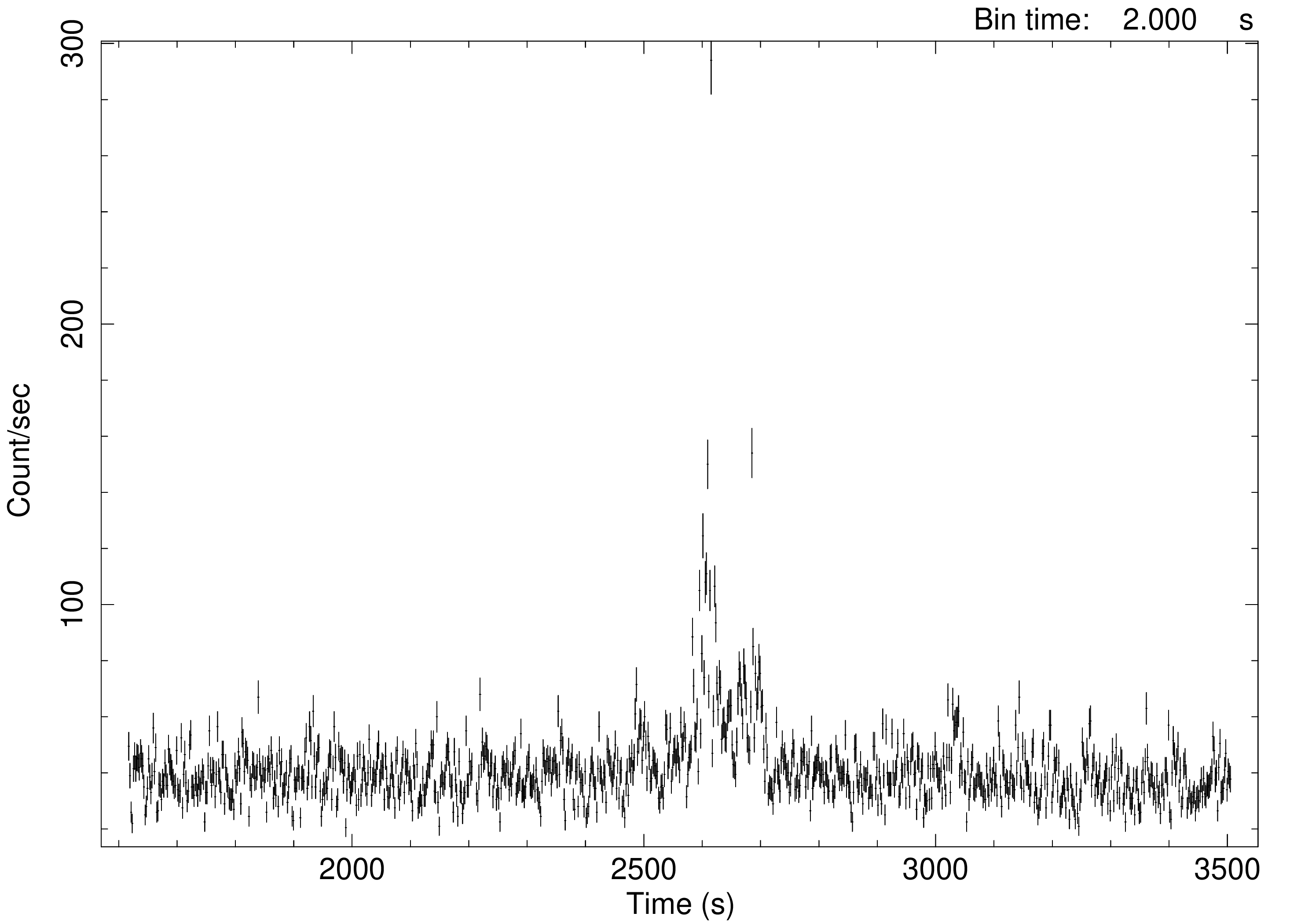}
    \caption{NICER light curve of the bursts from GRO J1744-28 during the first 2 ks of observation 4202210101 (2021-06-09) with a bin time of 2 s.}
    \label{fig:lc}
\end{figure}

The Neutron Star Interior Composition Explorer (NICER) X-Ray Timing Instrument (XTI), as a payload on board the International Space Station \citep{2016SPIE.9905E..1HG}, operates in the 0.2–12.0 keV range. NICER/XTI conducted a total of 17 observations of GRO J1744-28 during the 2021 outburst. In this study, we selected eight observations between 2021 June 9 and 2021 June 22, shown in Fig. \cref{fig:counts} with a net exposure time greater than 2000 seconds to obtain enough statistics. Table \cref{tab:ObsIDs} lists the details of the observations. The data reduction and analysis for these NICER observations of GRO J1744-28 were conducted using the NICER Data Analysis Software (NICERDAS) in \textsc{HEASoft} (v6.36), following the standard protocols recommended by the instrument team. The level 1 event files were processed with the \texttt{nicerl2} task to apply the latest calibration (CALDB) and to filter the data using standard good time intervals (GTIs). These criteria excluded periods of the passages through the South Atlantic Anomaly (SAA) and included the times when source elevation was $> 15^\circ$ and the Earth limb was $> 30^\circ$.

The source spectra were then extracted using \texttt{nicerl3-spect}. The recommended NICER systematics were automatically included. The background file was estimated and subtracted using the NICER's newest background model, called SCORPEON \citep{2024HEAD...2110536M}. The corresponding instrument response files (RMF and ARF) were generated using the \texttt{nicerrmf} and \texttt{nicerarf} tasks. For timing analysis, the barycentric corrections were applied to those events using the \texttt{barycorr} task, and the photon arrival time due to binary orbital motion was corrected using the orbital parameters from Fermi/GBM\footnote{\url{https://gammaray.nsstc.nasa.gov/gbm/science/pulsars/lightcurves/groj1744.html}}. Light curves were extracted by \texttt{nicerl3-lc} with a time resolution of 0.01 s. All extracted spectra were grouped to a minimum of 25 counts per bin using \texttt{ftgrouppha} tool. The spectral analyses in this paper were performed with XSPEC 12.13.1\citep{1996ASPC..101...17A}, adopting the chi-squared statistic option. Uncertainties of all parameters were calculated at the 68\% confidence level using the Markov chain Monte Carlo (MCMC) method with a chain length of 20 000 steps unless otherwise noted.

\begin{figure}
    \centering
    \includegraphics[width=.5\textwidth]{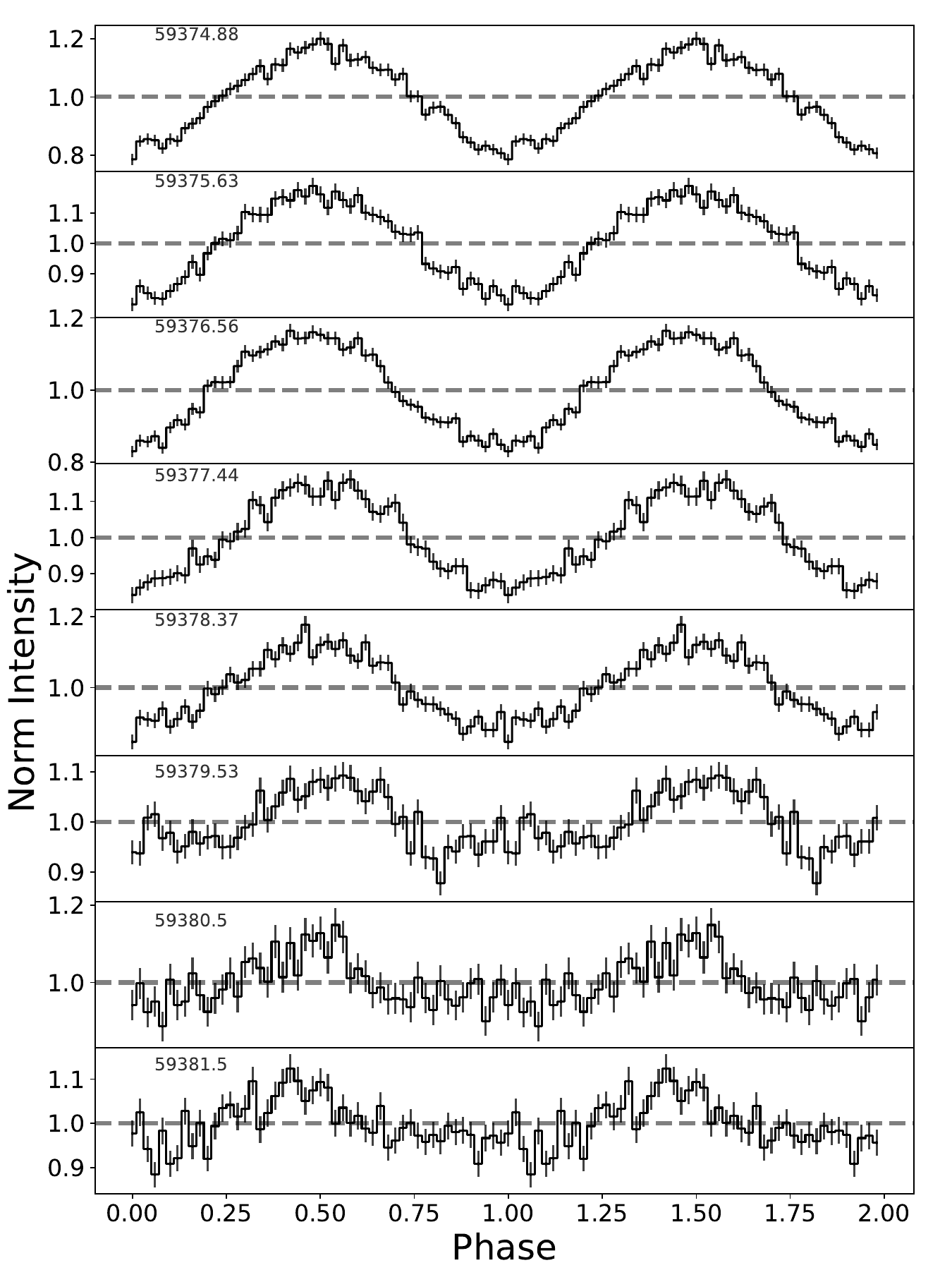}
    \caption{Pulse profiles in the 0.6–12 keV energy range from NICER at different epochs (labeled in MJD, left in each panel) during the 2021 outburst.}
    \label{fig:pulse_HE}
\end{figure}

\section{Results} \label{sec:results}

\subsection{Light curve and pulse profiles}

Figure \ref{fig:lc} shows the light curve during the first 2~ks (kiloseconds) of the NICER observation on 9 June 2021. Two type-II bursts are clearly visible, each with a typical peak duration of tens of seconds. The bursts were detected in five of the observations; at their peak, the count rate increased by approximately an order of magnitude relative to the persistent emission. In this work, we focus on the persistent emission and spectrum of GRO~J1744--28 by excluding the burst intervals from each observation.

We determined the pulse period using the epoch‑folding technique via the \textit{efsearch} task within the FTOOLS software package\footnote{A General Package of Software to Manipulate FITS Files, available at \url{https://heasarc.gsfc.nasa.gov/ftools}.}. Light curves in the 0.6–12.0 keV band were folded on a grid of trial periods around the initial spin period of 2.14 Hz, with 50 phase bins per trial period. The best‑fit period of 2.141128(7) Hz was identified by maximizing the $\chi^{2}$ statistic of the folded profile. The pulse profiles were then generated by folding each light curve at this determined period. As shown in Fig. \cref{fig:pulse_HE}, the profiles from all eight observations exhibit a single‑peaked shape.

\begin{figure}
    \centering
   \includegraphics[width=.24\textwidth]{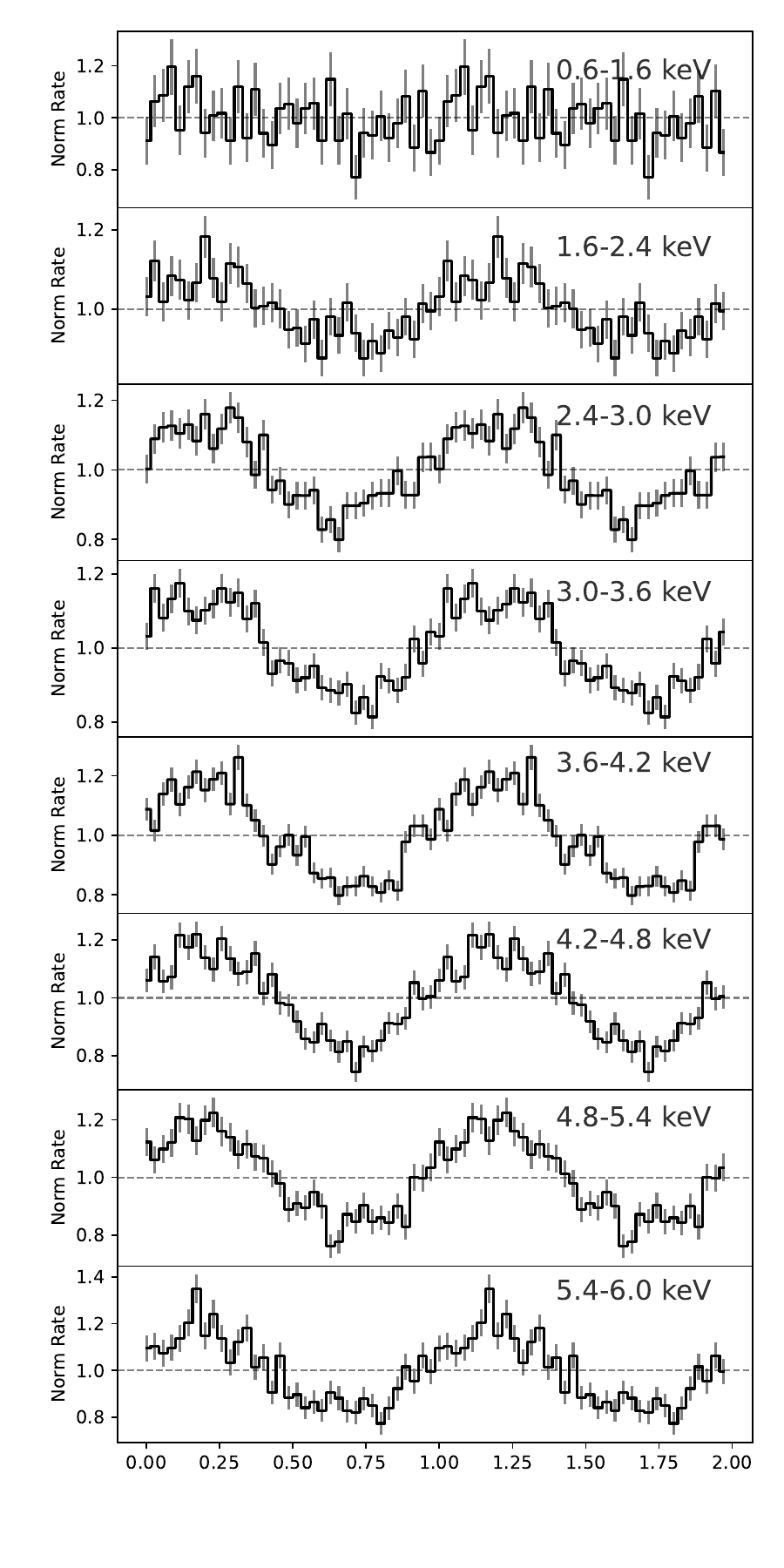}
   \includegraphics[width=.24\textwidth]{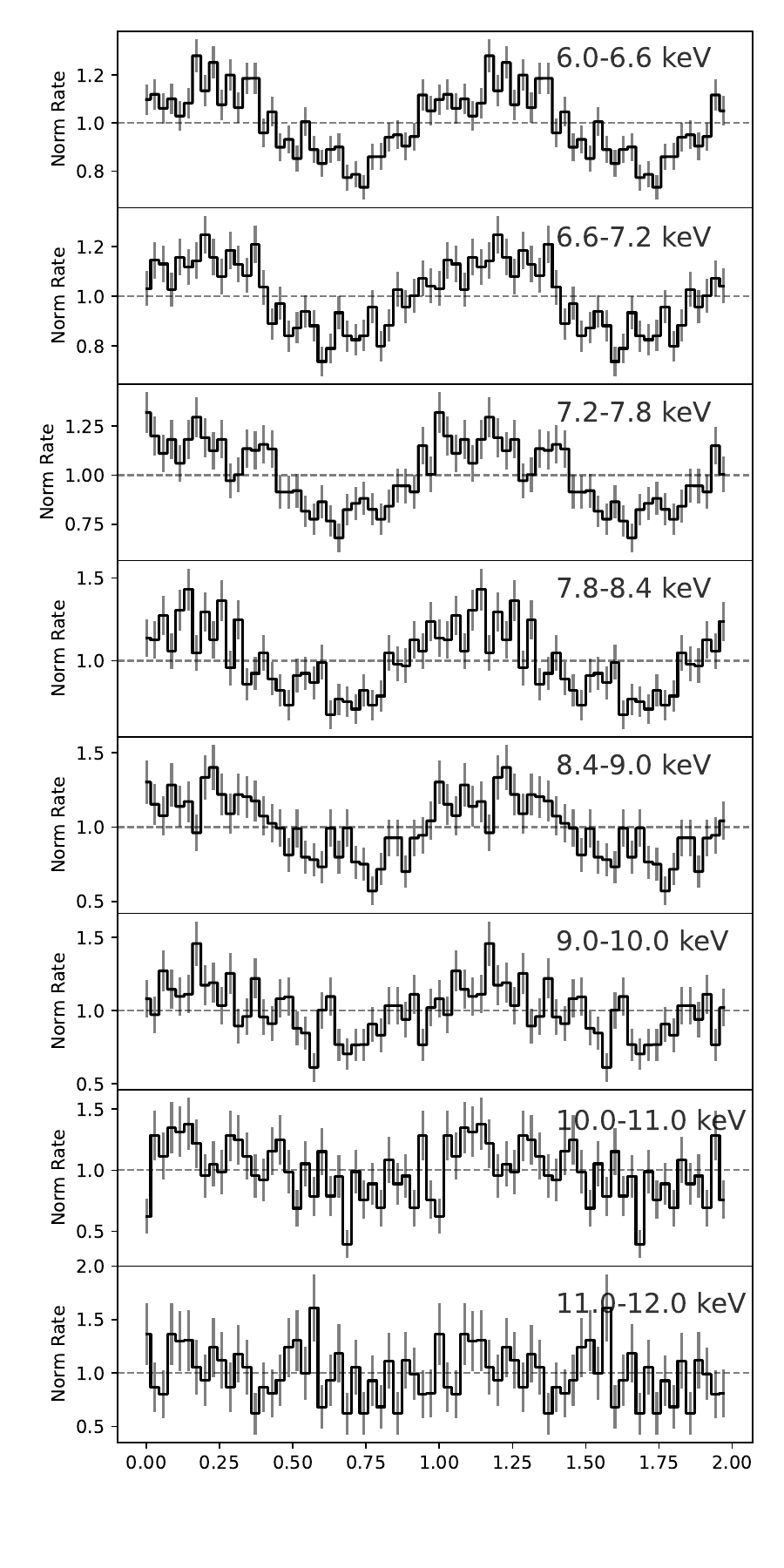}
    \caption{Pulse profiles of GRO~J1744--28 in different energy bands for the ObsID 4202210103.}
    \label{fig:pulse_profile}
\end{figure}

\begin{figure}
    \centering
    \includegraphics[width=.5\textwidth]{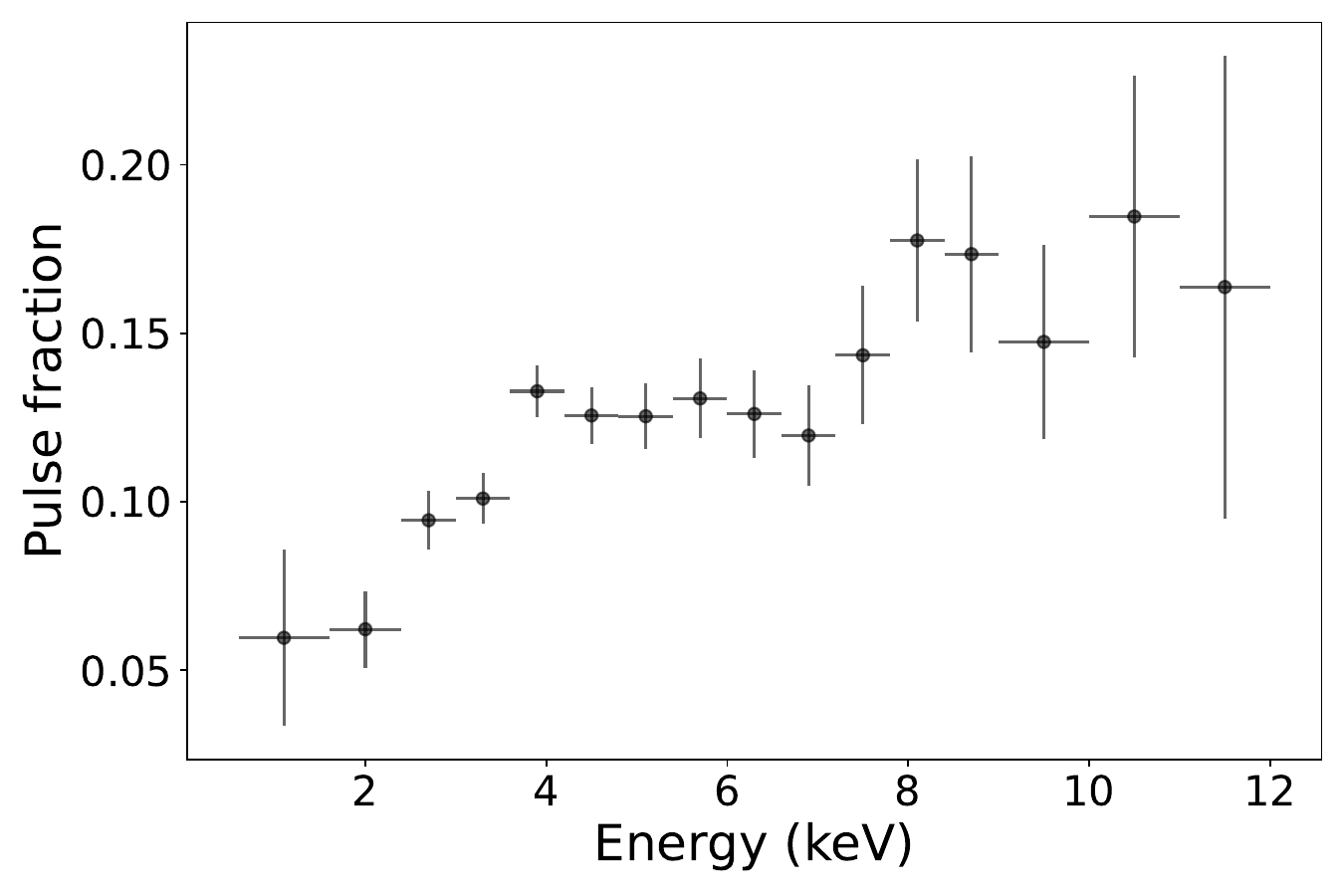}
    \caption{Pulse fraction as a function of energy for the ObsID 4202210103.}
    \label{fig:pf_energy}
\end{figure}

\begin{figure*}
    \centering
   \includegraphics[width=.49\textwidth]{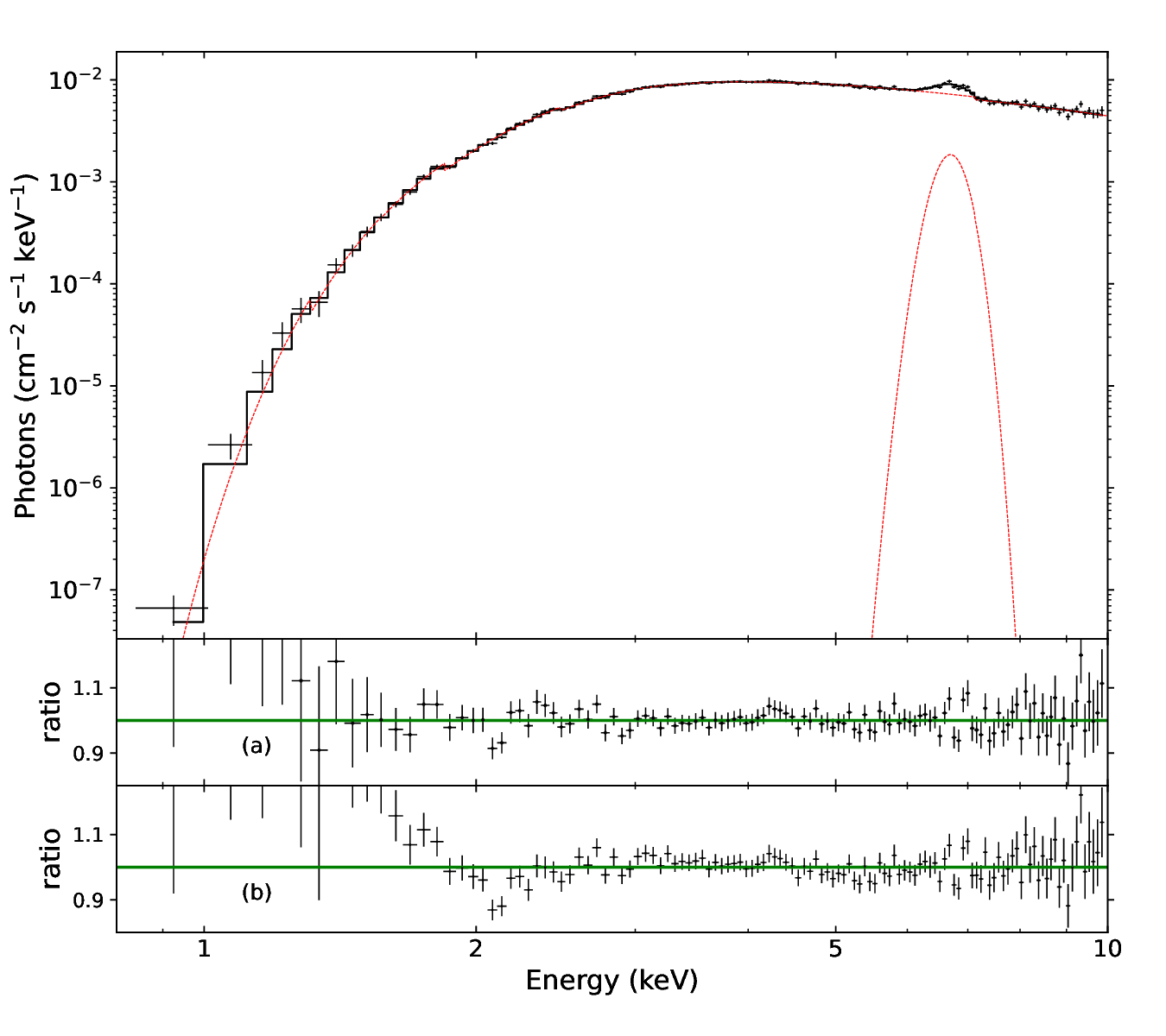}
   \includegraphics[width=.49\textwidth]{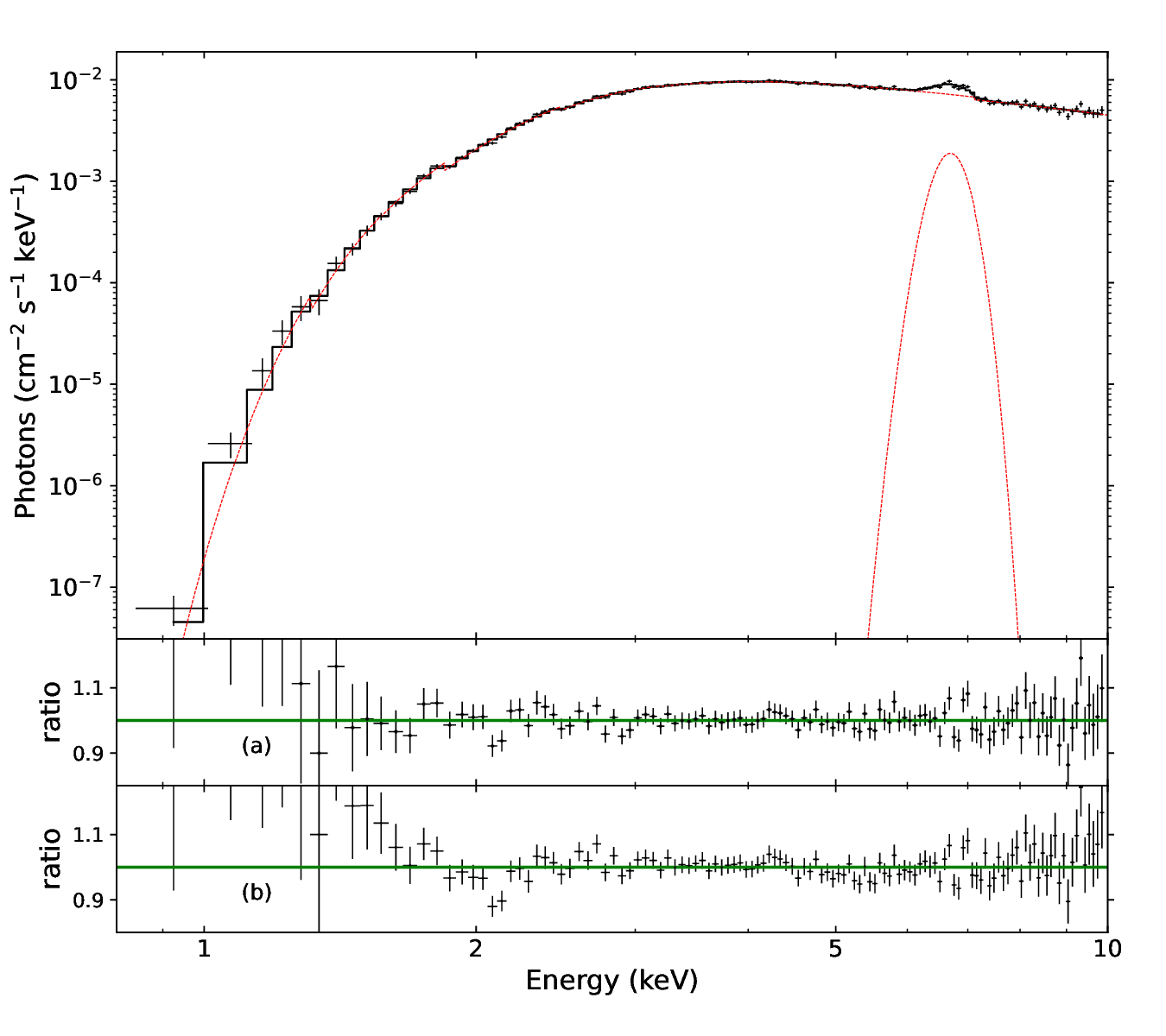}
   \includegraphics[width=.49\textwidth]{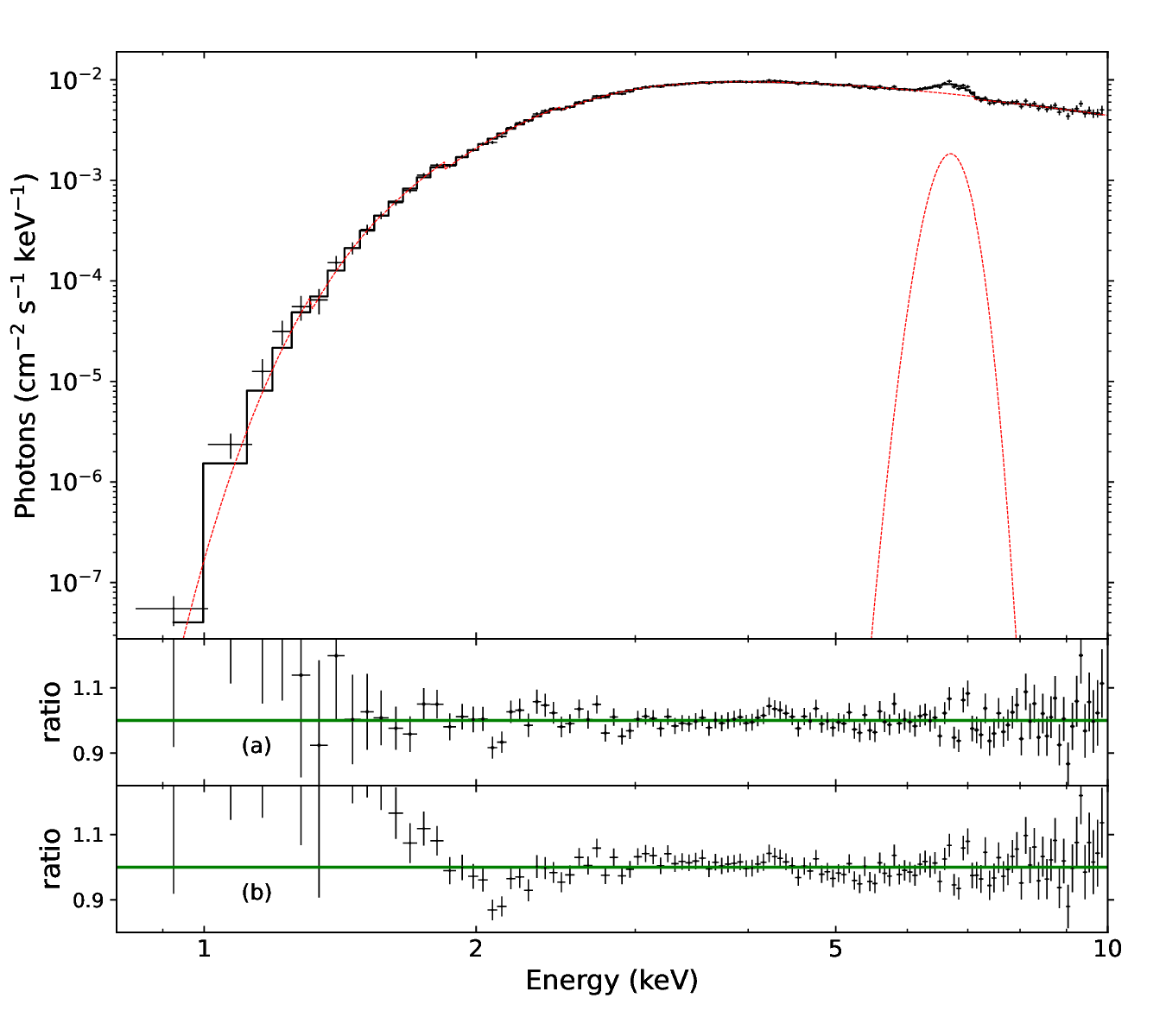}
   \includegraphics[width=.49\textwidth]{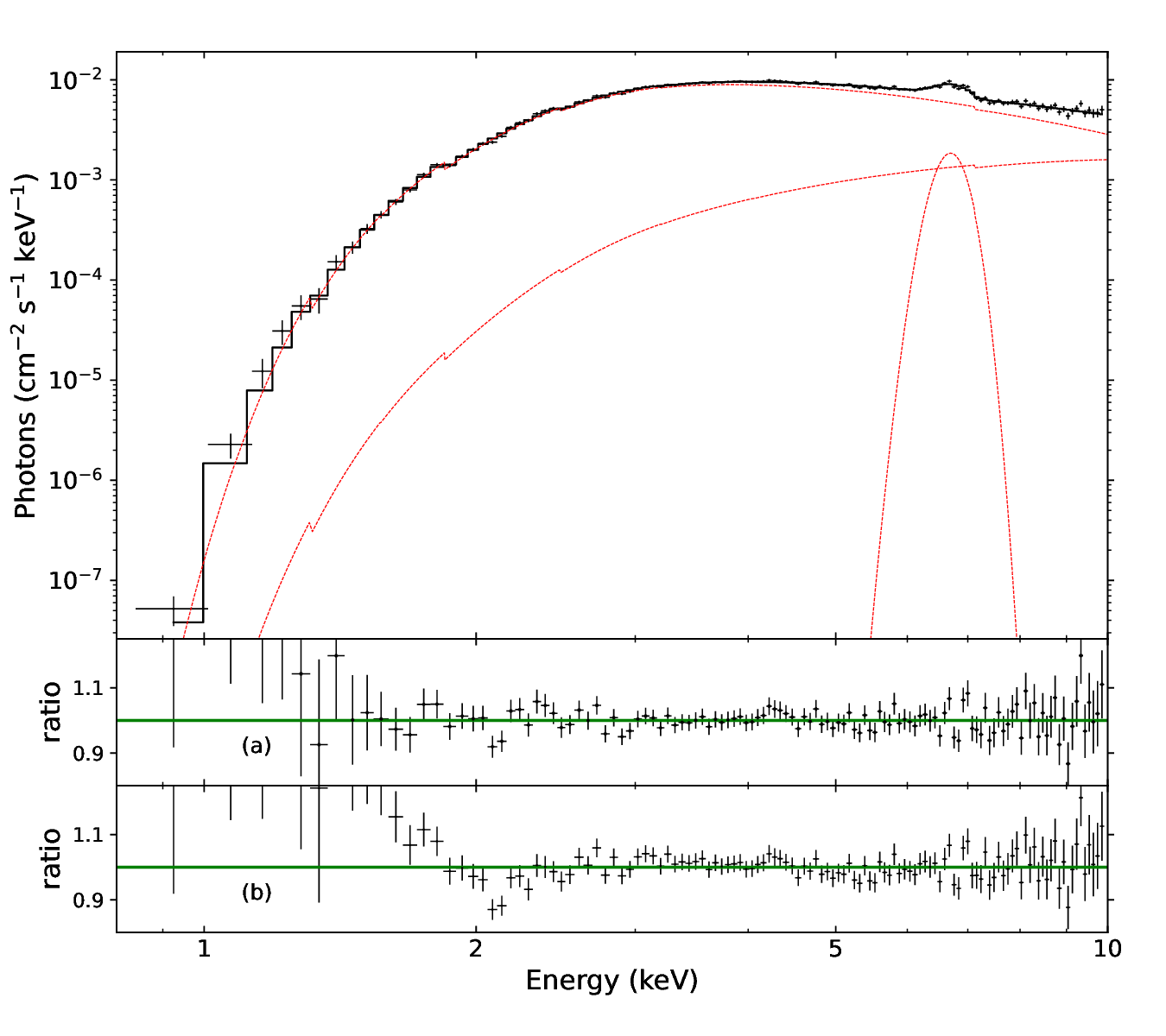}
    \caption{0.8--10~keV X-ray spectrum of GRO~J1744--28 (ObsID~4202210103) fitted with four phenomenological continuum models: cut-off PL (upper left), HIGHECUT (upper right), FDCUT (lower left), and NPEX (lower right). Panel (a): Residuals, data-to-model ratio, for the best-fitting model including a CRSF. Panel (b): Ratio for the best-fit model without a CRSF component.}
    \label{fig:spectrum}
\end{figure*}

We further examined the energy‑resolved pulse profiles using ObsID 4202210103, which provides the highest photon statistics. The light curve was divided into 16 contiguous energy bands and folded at the best‑fit pulse period. The resulting energy‑dependent pulse profiles are shown in Fig. ~\ref{fig:pulse_profile}. Clear, single‑peaked pulsations are detected across the 1.6–10 keV range. The lowest energy band (0.6–1.6 keV) and high band (10.0-12.0 keV) show no clear pulse profile, likely due to limited statistics. The root-mean-square (RMS) pulse fraction, defined as $PF_{\text{rms}} = \frac{1}{\bar{r}} \left( \frac{1}{N} \sum_{i=1}^{N} [(r_i - \bar{r})^{2}-\sigma_{r_i}^2] \right)^{1/2}$ \citep{2023A&A...677A.103F}, was calculated for each energy band, where $r_i$ is the count rate in the $i$-th phase bin, $\sigma_{r_i}$ is its uncertainty, $\bar{r}$ is the mean count rate, and $N = 35$ is the total number of phase bins. The uncertainty in $PF_{\text{rms}}$ was estimated via standard error propagation and is quoted at the 90\% confidence level. Its energy dependence is presented in Fig.~\ref{fig:pf_energy}, consistent with the energy dependence of the simple definition of the pulse fraction, $(I_{\text{max}}-I_{\text{min}})/(I_{\text{max}}+I_{\text{min}})$. In general, the pulse fraction increases with energy from 0.6 to 12 keV, except at $\sim$6.7 keV where it close to the fluorescent iron line. The dip feature around 2 keV is not observed due to limited accuracy, as reported by \cite{2015MNRAS.449.4288D,2015MNRAS.452.2490D}. This dip may hint at the presence of a cyclotron feature. 

\begin{table*}
\centering
\caption{Best-fitting parameters in different models.}

\large
\renewcommand\arraystretch{1.23}
\setlength{\tabcolsep}{6.5mm}{
\begin{tabular}{l|rrrr}
\hline
& \multirow{2}{*}{cut-off PL} & \multirow{2}{*}{HIGHECUT} & \multirow{2}{*}{FDCUT} & \multirow{2}{*}{NPEX} \\ \\

\hline
$N_{H}$  & $6.99_{-0.28}^{+0.42}$ & $6.98_{-0.29}^{+0.42}$ &$6.99_{-0.19}^{+0.35}$ &  $6.94_{-0.35}^{+0.21}$\\
$\Gamma$ & $0.54_{-0.03}^{+0.09}$  &   $0.76_{-0.03}^{+0.08}$    &  $0.66_{-0.04}^{+0.05}$ &  $0.25_{-0.06}^{+0.06}$ \\
$E_{\rm fold}$(keV) &     $11.25_{-0.69}^{+1.93}$ &  $17.54_{-1.34}^{+4.25}$ &     $9.07_{-1.15}^{+1.11}$   &   $5.06_{-0.32}^{+0.40}$ \\
$E_{\rm cut}$(keV) &  ... &   $4.18_{-0.22}^{+0.16}$  &     $1.74_{-0.16}^{+0.11}$   &  ...  \\ \hline
$E_{\rm Fe}$(keV) & $6.70_{-0.04}^{+0.03}$  & $6.70_{-0.01}^{+0.02}$  &  $6.70_{-0.03}^{+0.02}$  &  $6.70_{-0.04}^{+0.04}$ \\
$\sigma_{\rm Fe}$(keV) &  $0.26_{-0.03}^{+0.04}$ & $0.26_{-0.01}^{+0.03}$  &   $0.26_{-0.01}^{+0.02}$   &   $0.26_{-0.02}^{+0.05}$ \\
$I_{\rm Fe}$ & $0.0016_{-0.0002}^{+0.0002}$ & $0.0016_{-0.0001}^{+0.0002}$  &  $0.0016_{-0.0001}^{+0.0001}$  & $0.0016_{-0.0002}^{+0.0002}$ \\ \hline
 $E_{\rm cyc}$ (keV) &   $1.77_{-0.22}^{+0.08}$ &   $1.80_{-0.17}^{+0.10}$  &  $1.74_{-0.16}^{+0.11}$ &     $1.82_{-0.15}^{+0.10}$  \\
 $\sigma_{\rm cyc}$ (keV) &   $0.55_{-0.05}^{+0.14}$ & $0.49_{-0.07}^{+0.12}$ & $0.58_{-0.06}^{+0.08}$   &  $0.52_{-0.07}^{+0.11}$ \\
 $d_{\rm cyc}$ (keV) &  $0.53_{-0.13}^{+0.38}$ &   $0.45_{-0.14}^{+0.27}$  &  $0.60_{-0.13}^{+0.37}$ &   $0.44_{-0.13}^{+0.25}$\\ \hline
$\chi^2_{\rm red}$ (dof) no CRSF & 1.53 (121) & 1.25 (120) & 1.56 (120)   &  1.49 (120) \\ 
$\chi^2_{\rm red}$ (dof) & 0.95 (118) & 0.90 (117) & 0.96 (117)   &  0.95 (117) \\ \hline
$\rm Flux_{0.8-10}$ & {$0.710_{-0.002}^{+0.001}$} &  {$0.712_{-0.002}^{+0.001}$} & {$0.710_{-0.002}^{+0.001}$} & {$0.706_{-0.001}^{+0.001}$}  \\
\hline
\end{tabular}
}
%\begin{tablenotes}
\tablefoot{
The flux is given in units of $10^{-9}$ erg cm$^{-2}$ s$^{-1}$, and the column density $N_{\rm H}$ is listed in units of 10$^{22}$ atoms cm$^{-2}$. \\
%\end{tablenotes}
}
\label{tab:pars}
\end{table*}

\begin{figure}
    \centering
   \includegraphics[width=.49\textwidth]{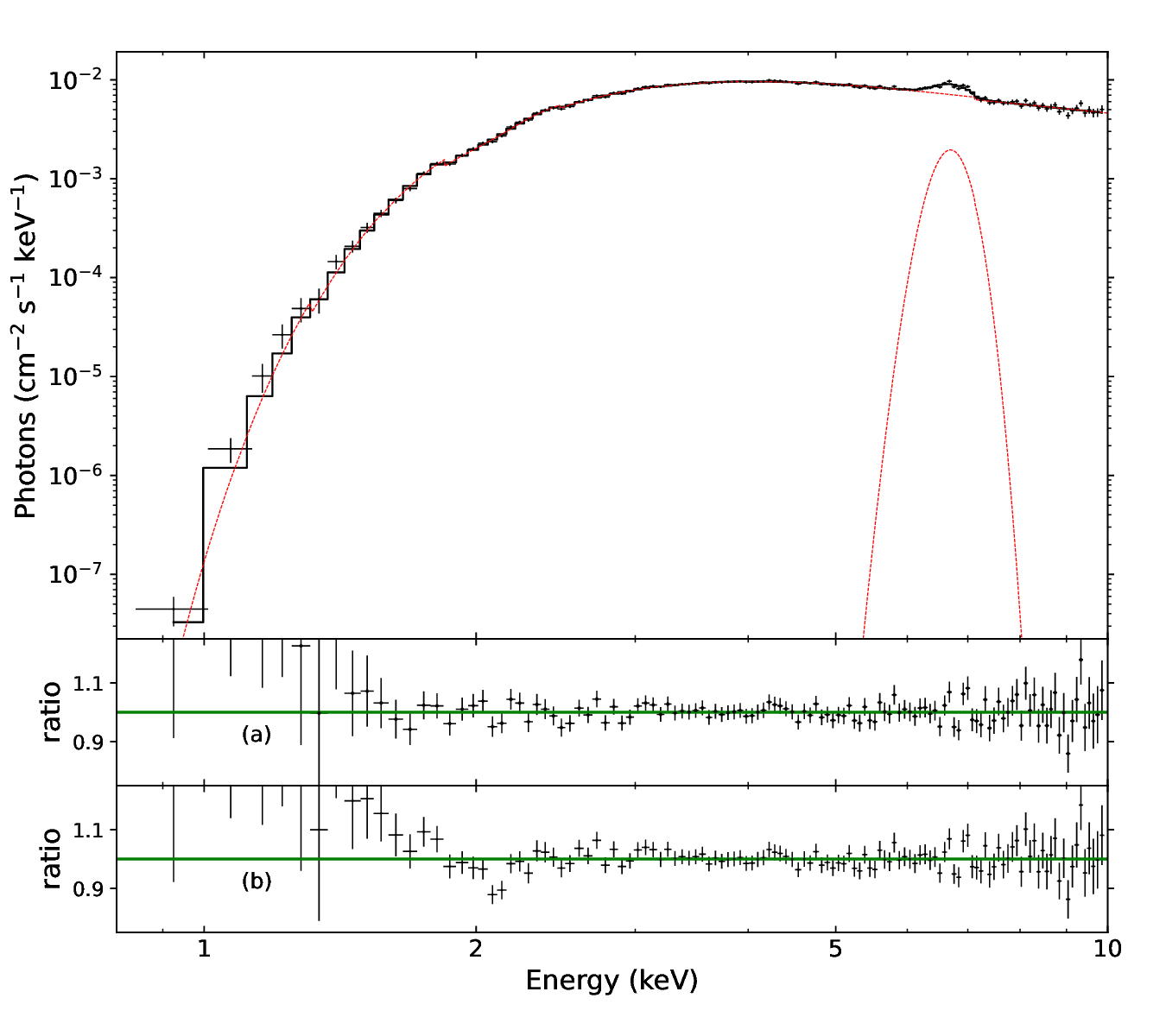}
    \caption{Spectral analysis of GRO~J1744--28 in the 0.2--10~keV band. Panel~(a): Ratio of the data to the model for the best-fit CompTT continuum with an additional Gaussian absorption line (gabs). Panel~(b): Ratio of the data to the model for the best-fit CompTT continuum without the gabs component.}
    \label{fig:comptt}
\end{figure}

\begin{table*}[htp]
\centering
\caption{Spectral parameters for the CompTT model.}
\label{tab:comptt}

\renewcommand\arraystretch{1.5}
\setlength{\tabcolsep}{0.5mm}{
\begin{tabular}{l|cccccccccccc}
\hline
  & $N_{\rm H}$ &  $T0$ & $kT_e$  & $\tau_{\rm e}$ & $E_{\rm Fe}$&  $\sigma_{\rm Fe}$&  $I_{\rm Fe}$&  $E_{\rm cyc}$&  $\sigma_{\rm cyc}$&  $d_{\rm cyc}$ & Flux & $\chi^2$/dof \\ 
\hline 
  4202210103 & $6.70_{-0.13}^{+0.13}$ & $0.93_{-0.03}^{+0.02}$ & $11.10_{-1.29}^{+1.71}$ &  $5.08_{-0.17}^{+0.15}$ & $6.69_{-0.02}^{+0.03}$ & $0.28_{-0.02}^{+0.04}$ & $0.0015_{-0.0002}^{+0.0001}$ &  &   &  & $0.679_{-0.002}^{+0.002}$ &  142/120 \\ \hline
 & $6.38_{-0.30}^{+0.31}$ & $0.94_{-0.05}^{+0.04}$ & $11.60_{-1.40}^{+1.65}$ &  $5.11_{-0.21}^{+0.22}$ & $6.70_{-0.05}^{+0.03}$ & $0.28_{-0.04}^{+0.08}$ & $0.0015_{-0.0002}^{+0.0003}$ & $2.07_{-0.03}^{+0.05}$ & $0.15_{-0.08}^{+0.12}$  & $0.05_{-0.03}^{+0.04}$ & $0.670_{-0.002}^{+0.002}$ &  105/117 \\ \hline
\end{tabular} 
}
\end{table*}

\subsection{Spectral properties}

\subsubsection{Phase-averaged spectroscopy} \label{sec:averaged}

The X-ray spectra of GRO J1744-28 are typically characterized by a power law modified by an exponential cut-off, along with the iron emission lines. For the spectral analysis, we fitted the data with four commonly applied empirical continuum models in accreting pulsars, the cut-off power law (cut-off PL), the absorbed power law with a high energy cut-off (HIGHECUT), the Fermi–Dirac cut-off power law (FDCUT), and the negative positive exponential (NPEX) model. Additionally, interstellar absorption was included by the tbabs model in XSPEC, and we adopted the {\tt wilm} abundances \citep{2000ApJ...542..914W}. First, we applied a simple cut-off PL ($cutoffpl$ in XSPEC) to fit the spectra in GRO J1744-28 \citep{2015ApJ...804...43Y,2020A&A...643A.128K}, which is characterized as 
\begin{equation}
I(E)=N \cdot E^{-\Gamma}\left[\exp \left(\frac{E}{E_{\text {fold }}}\right)\right]^{-1}.
\end{equation}
Here, $\Gamma$ is the power-law index, and $E_{\text {fold }}$ is the folding energy. A general continuum spectrum of HIGHECUT ($pow * highecut$ in XSPEC) is typically used to fit the spectra, as applied to GRO J1744-28 \citep{2015MNRAS.452.2490D}. This is characterized by
\begin{equation}
I(E)=\left\{\begin{array}{ll}
N E^{-\Gamma} & \text { for } E \leq E_{\text {cut }}, \\
N E^{-\Gamma} \times \exp \left[-\left(E-E_{\text {cut }}\right) / E_{\text {fold }}\right] & \text { for } E>E_{\text {cut }} ,
\end{array}\right.
\end{equation}
where $E_{\text {cut }}$ is the cut-off energy. Another form of power-law-like continuum, the FDCUT model, was also employed to fit the continuum. It is expressed by
\begin{equation}
I(E)=N \cdot E^{-\Gamma}\left[1+\exp \left(\frac{E-E_{\text {cut }}}{E_{\text {fold }}}\right)\right]^{-1}.
\end{equation}
Finally, the NPEX model, a sum of a positive and negative power law multiplied by an exponential cut-off, was also previously used \citep{Mihara1995}. This is expressed by
\begin{equation}
I(E)=\left(N_1 \cdot E^{-\Gamma_1}+N_2 \cdot E^{+\Gamma_2}\right) \exp \left(-E / E_{\mathrm{fold}}\right),
\end{equation}
where $\Gamma_2$ is usually set to 2. As mentioned, the CRSFs have been reported by different missions previously. To model the CRSF, we used a multiplicative absorption line model with a Gaussian optical depth profile (gabs in XSPEC):
\begin{equation}
gabs(E)=\exp \left(-\frac{d_{\mathrm{cyc}}}{\sqrt{2 \pi} \sigma_{\mathrm{cyc}}} e^{-0.5 \left[\left(E-E_{\mathrm{cyc}}\right) / \sigma_{\mathrm{cyc}}\right]^2}\right) ,
\end{equation}
where $E_{\mathrm{cyc}}$ is the cyclotron line energy, with a line depth $d_{\mathrm{cyc}}$ and a width $\sigma_{\mathrm{cyc}}$. The iron line emission feature around 6.7 keV (Gaussian in XSPEC) was also added to our fitting. 

\begin{figure}
    \centering
    \includegraphics[width=.49\textwidth]{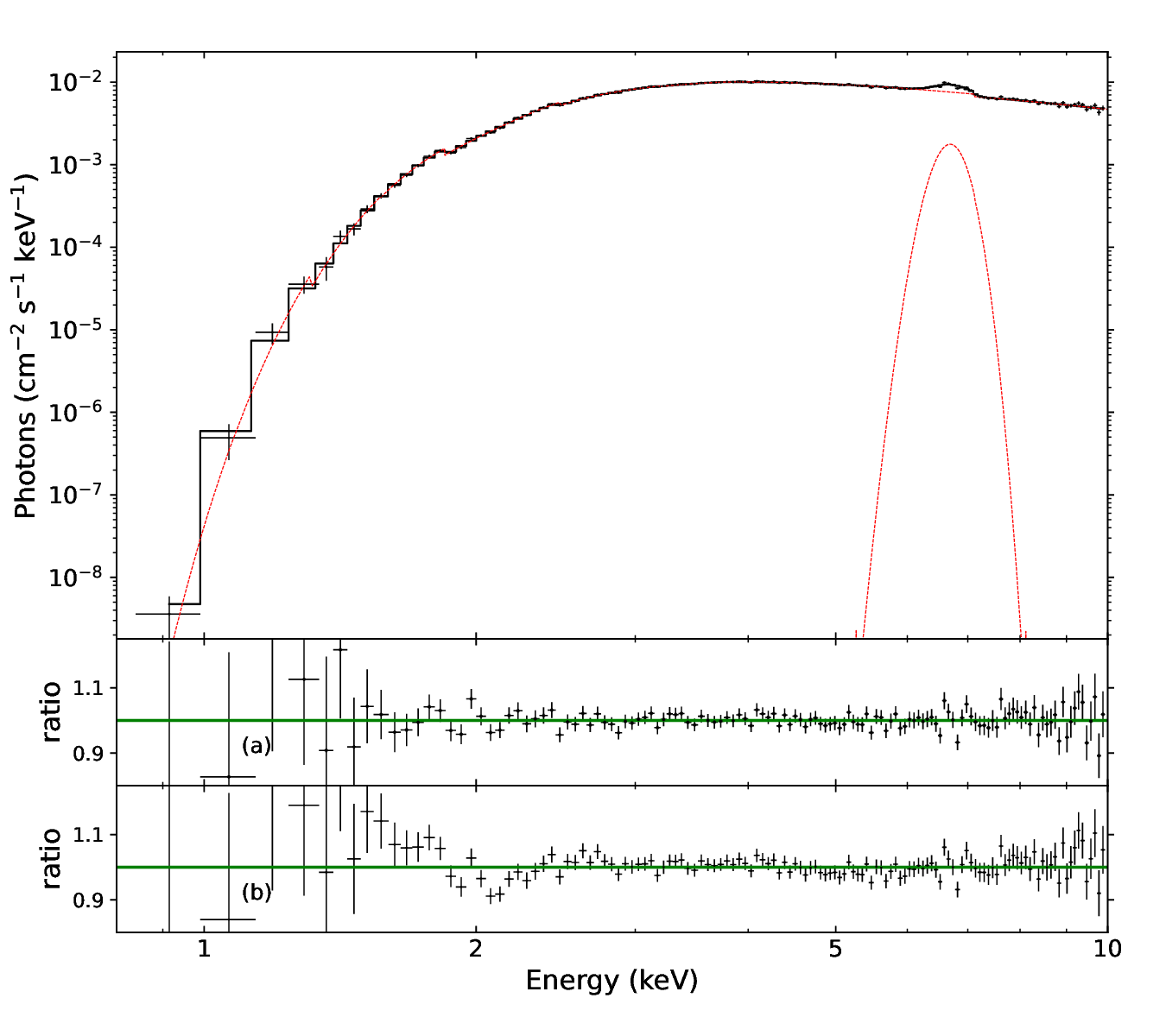}
    \includegraphics[width=.49\textwidth]{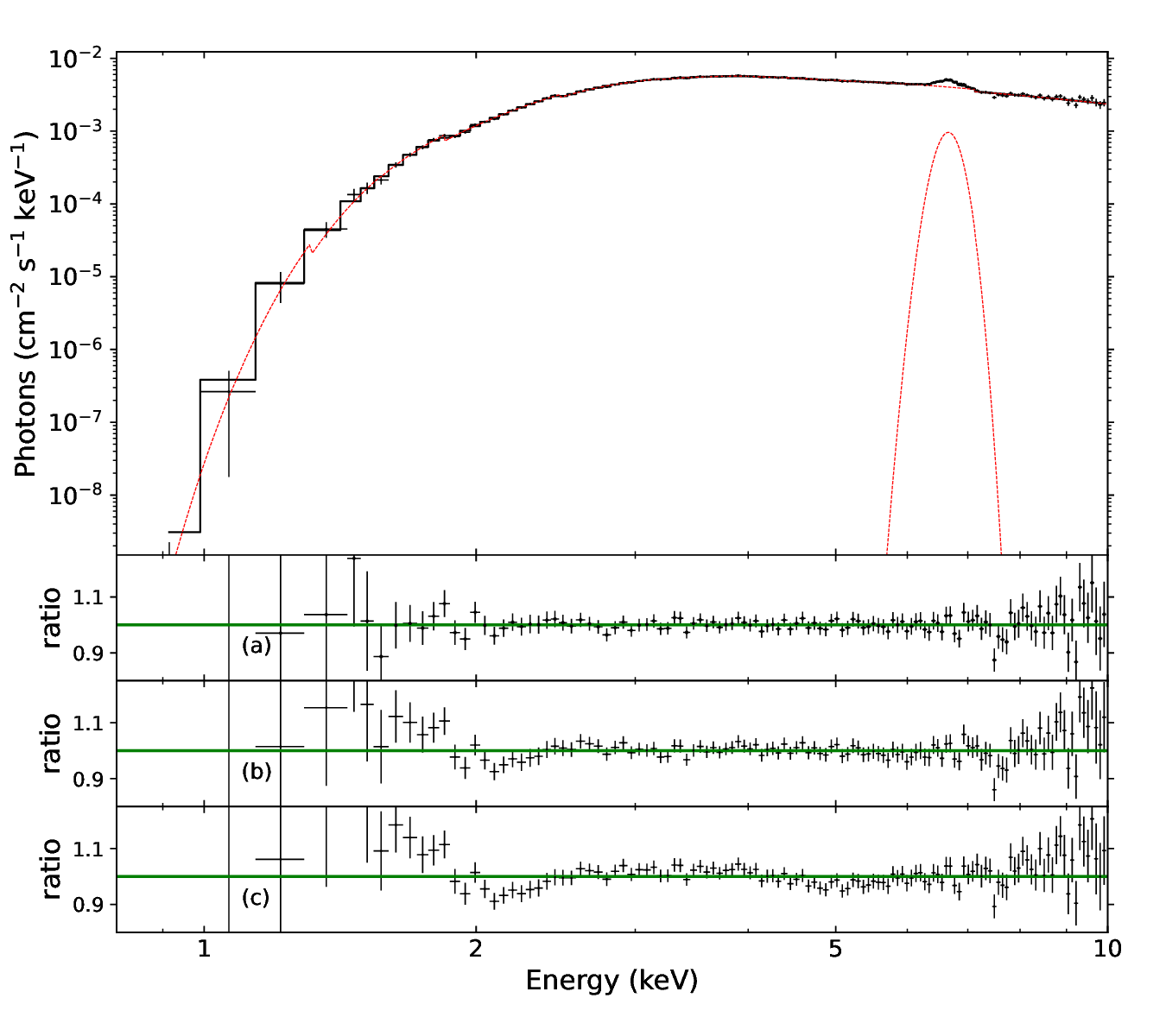}
    \caption{Spectral fitting results for GRO~J1744--28 in the 0.8--10~keV band using the HIGHECUT model. Top panel: Observation 1. Subpanel (a): Ratio for the best-fitting model including a CRSF. Subpanel (b): Ratio for the best-fit model without a CRSF component. Bottom panel: Observation 2. Subpanel (a): Ratio for the overall best-fitting model. Subpanel (b): Ratio for the best-fit model including the 5~keV CRSF. Subpanel (c): Ratio for the best-fit model without the two CRSF components.}
    \label{fig:spec_merged}
\end{figure}

\begin{table*}[htp]
\centering
\caption{Best-fitting spectral parameters of GRO J1744-28 in 0.8–10 keV during the 2021 outburst with NICER.}
\label{tab:ObsIDs_spec}

\renewcommand\arraystretch{1.5}
\setlength{\tabcolsep}{0.5mm}{
\begin{tabular}{l|cccccccccccc}
\hline
  Observation  & $N_{\rm H}$ & $\Gamma$ & $E_{\rm cut}$  & $E_{\rm fold}$ & $E_{\rm cyc}$&  $\sigma_{\rm cyc}$&  $d_{\rm cyc}$&  $E_{\rm cyc2}$&  $\sigma_{\rm cyc2}$&  $d_{\rm cyc2}$ & Flux & $\chi^2$/dof \\ 
\hline 
1   & $7.96_{-0.31}^{+0.13}$ & $0.77_{-0.05}^{+0.06}$ & $3.72_{-0.18}^{+0.20}$ &  $16.62_{-1.21}^{+3.61}$ & $2.06_{-0.08}^{+0.02}$ & $0.25_{-0.03}^{+0.11}$ & $0.10_{-0.02}^{+0.08}$ &  &   &  & $0.790_{-0.001}^{+0.001}$ &  110/129 \\ \hline
2  & $7.99_{-0.28}^{+0.26}$ & $0.92_{-0.05}^{+0.06}$  & $3.86_{-0.31}^{+0.22}$ & $18.46_{-2.22}^{+2.94}$ & $2.02_{-0.12}^{+0.08}$ & $0.39_{-0.05}^{+0.10}$ & $0.24_{-0.07}^{+0.17}$ & $5.03_{-0.14}^{+0.25}$ & $0.52_{-0.14}^{+0.43}$ & $0.05_{-0.03}^{+0.04}$ & $0.431_{-0.001}^{+0.001}$ & 80/121  \\ \hline
\end{tabular} 
}

\tablefoot{
The unabsorbed flux is given in units of $10^{-9}$ erg cm$^{-2}$ s$^{-1}$, and the column density $N_{\rm H}$ is listed in units of 10$^{22}$ atoms cm$^{-2}$ . The parameters $E_{\rm cut}$ (cut-off energy), $E_{\rm fold}$ (the folding energy), $E_{\rm cyc}$, $\sigma_{\rm cyc}$, and $d_{\rm cyc}$ are in units of keV. Uncertainties are given at the 68\% confidence level. \\
}
\end{table*}

As an example, we present the fitting results for ObsID 4202210103 using the four continuum models. The best-fit spectra are shown in Fig.~\ref{fig:spectrum}, and the corresponding parameters are listed in Table~\ref{tab:pars}. We find that all models describe the spectrum adequately, yielding acceptable reduced chi-squared values ($\chi^{2}_{\mathrm{red}}$). As shown in panel~(b) of Fig.~\ref{fig:spectrum}, clear residuals around 2~keV are present for all models---the cut-off PL, HIGHECUT, FDCUT, and NPEX ($\chi^{2}_{\mathrm{red}}$ values ranging from 1.25 to 1.56; see Table~\ref{tab:pars}). This absorption-like feature near 2~keV may indicate an additional CRSF component, which we modelled with a multiplicative Gaussian absorption profile (\texttt{gabs} in XSPEC). Including this component significantly improved the fits, as seen in the residuals in panel~(a) of Fig.~\ref{fig:spectrum}. The resulting line parameters are listed in Table~\ref{tab:pars}. Notably, the cyclotron line parameters are consistent across the different continuum models, indicating that the choice of continuum model does not significantly affect the line energy. We also tested modelling the line with a Lorentzian profile (\texttt{CYCLABS} in XSPEC), which yielded comparable values. For example, the energy, width, and depth parameters are found to be at 
$1.86_{-0.10}^{+0.07}$\,keV, $0.48_{-0.05}^{+0.06}$\,keV, $0.40_{-0.06}^{+0.11}$ 
for the cut-off PL model, and 
$1.94_{-0.08}^{+0.05}$\,keV, $0.38_{-0.03}^{+0.09}$\,keV, $0.31_{-0.05}^{+0.08}$ 
for the HIGHECUT model. We also applied a physical CompTT model \citep{1994ApJ...434..570T} to describe the Comptonization of soft photons, characterized by the soft photon temperature $T_0$, the plasma temperature $kT_{\mathrm{e}}$, and the plasma optical depth $\tau_{\mathrm{e}}$. The model provides a good fit to the spectrum, yielding a reduced $\chi^2$ of approximately 0.90. The spectral fit is presented in Fig.~\cref{fig:comptt}, where clear residuals are evident around 2~keV. The best-fit parameters are summarized in Table~\cref{tab:comptt}. For subsequent analysis, we adopted the empirical power law with a high-energy cut-off (HIGHECUT) plus gabs model for GRO~J1744--28, due to its better $\chi^{2}_{\mathrm{red}}$ for continuum description and fit stability. 

To improve the statistical quality and signal-to-noise ratio of the data, we merged the spectra from the eight observations into two groups using the task \textit{niobsmerge}: observation 1 (combining ObsID 4202210101 with 4202210104) and observation 2 (combining ObsID 4202210105 with 4202210108). For observation~1, as shown in the top panels of Fig.~\cref{fig:spec_merged}, the best-fit using the HIGHECUT continuum model yields a reduced $\chi^2$ of 1.30 for 132 degrees of freedom (d.o.f.). However, residuals are clearly seen around 2~keV in panel~(b). After adding a CRSF component, the fit improves substantially, resulting in a reduced $\chi^2$ of 0.85 for 129 d.o.f. (panel~(a)). The best-fit parameters for the CRSF in observation~1 are a centroid energy of $2.06_{-0.08}^{+0.02}$~keV, a width of $0.25_{-0.03}^{+0.11}$~keV, and a depth of $0.10_{-0.02}^{+0.08}$. To evaluate the significance of the cyclotron line candidate, we generated 10,000 simulated spectra using the \texttt{fakeit} command in XSPEC, adopting the same exposure time and spectral grouping as the real data. The simulation was based on the null-hypothesis model of the best-fit model without the absorption line component. Each simulated spectrum was first fitted with this null-hypothesis model and the resulting best-fit model was then used to again simulate the new spectrum. Finally, we fitted these simulated spectra using the best model with and without the line component. The difference in fit statistics, $\Delta \chi^2$, was calculated for each simulation to build its distribution. We find that the observed $\Delta \chi^2$ of 60.6 (when adding a CRSF line) is significantly greater than any value from the distribution. Assuming the $\Delta \chi^2$ distribution follows a $\chi^2$ distribution with three d.o.f. (as expected for three additional parameters), the chance probability is $ \sim 2.9 \times 10^{-13} $, implying a $ \sim 7.3\sigma $ detection. However, the simulations do not include these systematic uncertainties, especially in the calibration-sensitive 1.8-2.3 keV band, which can generate structured spectral residuals that mimic a broad absorption line. As a result, the reported statistical significance is probably overestimated.

For observation~2 during the 2021 outburst, we find that the residuals of the best-fit continuum model show an absorption feature at $\sim$4.5\,keV (see panel~(c) of the bottom plots in Fig.~\cref{fig:spec_merged}), as previously reported from \textit{BeppoSAX} observations in 1997 \citep{2015MNRAS.452.2490D} and from \textit{XMM-Newton} and \textit{INTEGRAL} observations in 2014 \citep{2015MNRAS.449.4288D}. We therefore first added a cyclotron line near 5\,keV to the model. This inclusion improves the fit, reducing the reduced $\chi^2$ from 1.35 for 127 d.o.f. to 0.97 for 124 d.o.f., with a null-hypothesis probability of 0.572. However, significant residuals around 2\,keV remain visible (panel~(b)). Including a second CRSF at this energy further improves the fit, yielding a reduced $\chi^2$/d.o.f. = 0.66/121 and a null-hypothesis probability of 0.999. The centroid energies of the two lines are $5.03_{-0.14}^{+0.25}$\,keV and $2.02_{-0.12}^{+0.08}$\,keV, respectively. The best-fit parameters for the two merged observations are listed in Table~\cref{tab:ObsIDs_spec}. We note that the 5\,keV cyclotron line is not detected in observation~1. This is consistent with the fact that the previously reported $\sim$4.7\,keV feature is not always present, which may depend on the accretion state or geometry. Another possibility for the absence of the 5 keV line in observation 1 is the narrow energy band of NICER: when applying a broadband model to a narrow energy band, the additional bending will flatten weak absorption features. 

To further investigate the $\sim$2.0$\,$keV feature and alternative origins such as an unmodelled soft excess or a detector feature (such as the silicon K edge), we performed additional spectral fits to observation~1. First, we tested for a soft excess by including a \texttt{diskbb} component. This did not improve the fit statistically ($\chi^2 > $1.3). Second, we modelled the feature with an absorption edge. While the resulting fit was acceptable, the observed $\Delta\chi^2 = 49.2$ was worse than the statistic improvement ($\Delta\chi^2 = $ 60.6) achieved by including a Gaussian absorption line (\texttt{gabs}). Furthermore, the inferred line width and depth from the \texttt{gabs} model are physically consistent with the expectations for a CRSF. We therefore conclude that the significant absorption feature at $\sim$2.0$\,$keV is a potential cyclotron line candidate. However, we note the instrumental features in the NICER response, which could produce, at least in part, the observed residuals. In the model without absorption line, the residuals reach up to $\sim$10\%, whereas the instrumental residual artefacts remain within 2-3\% , with larger features at the specific energies such as the silicon K edge at 1.84 keV and the Au M-edge near 2.2 keV \footnote{\url{https://heasarc.gsfc.nasa.gov/docs/nicer/analysis_threads/arf-rmf/}}. Thus, additional, high-quality observations are required to examine the robustness of this absorption feature in the future.

\begin{figure}
    \centering
    \includegraphics[width=.49\textwidth]{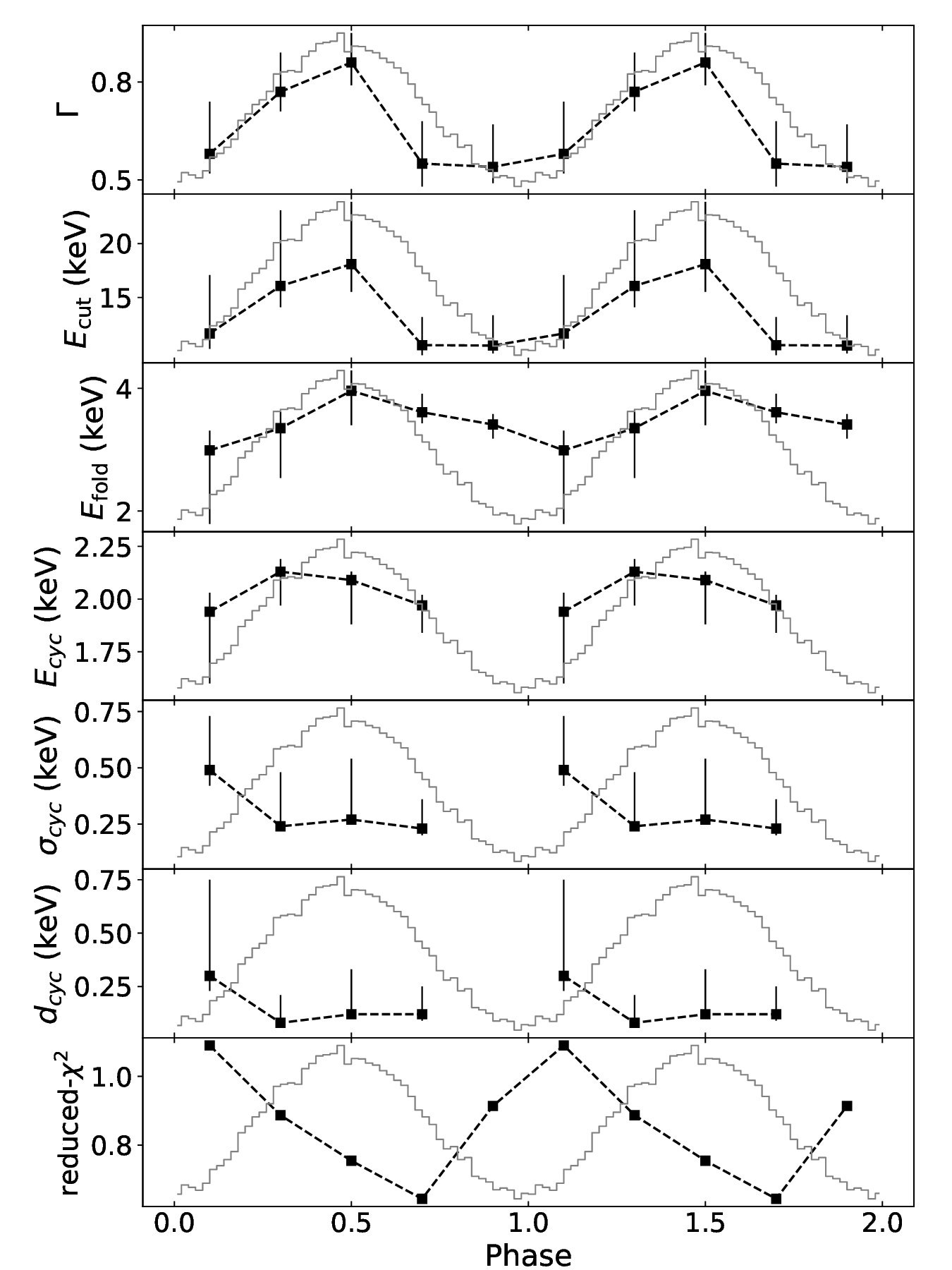}
    \caption{Cyclotron line parameters and spectral parameters shown over the pulse phases. The grey line represents the 0.6--12 keV pulse profile.}
    \label{fig:Par_Phase}
\end{figure}

\subsubsection{Phase-resolved spectroscopy} \label{sec:resolved}

In many accreting X-ray pulsars, the centroid energy of the cyclotron line varies with pulse phase, a behaviour linked to the accretion geometry and magnetic field configuration (e.g. \citealt{2019A&A...622A..61S}). To investigate such variations in GRO~J1744--28, we performed phase-resolved spectroscopy using observation 1 as a representative case since we mainly focused on the 2 keV cyclotron line. The pulse profile was divided into five phase bins, and for each bin, a spectrum and its corresponding response matrix were generated using the standard NICER pipeline. All phase-resolved spectra are well described by a HIGHECUT continuum modified by a Gaussian absorption line (\texttt{gabs} in XSPEC). The best-fitting spectral parameters for each phase bin, along with the 0.6--12~keV pulse profile, are shown in Fig.~\ref{fig:Par_Phase}. We find that the CRSF is significantly detected in the phase interval 0.0--0.8 (e.g. $\sim 4 \sigma$ for the 0.3 phase), which coincides with the peak of the pulse profile.

As shown in Fig.~\ref{fig:Par_Phase}, the centroid energy of the CRSF has a maximum near a pulse phase of 0.2, which corresponds to the ascending phases of the pulse profile. The line width $\sigma$ varies from 0.2 to 0.5~keV, while its depth $d$ ranges between 0.1 and 0.3 across the pulse. The photon index $\Gamma$ (average $\approx 0.7$) shifts from 0.5 to 0.9, with a maximum around the peak phase. Both the cut-off energy $E_\mathrm{cut}$ and the folding energy $E_\mathrm{fold}$ also show clear phase dependence. The CRSF is clearly detected in the peak-phase
bins of the pulse-resolved spectra, which also indicate evident phase dependence.

\section{Discussion and summary} \label{sec:discussion}

\begin{figure}
    \centering
    \includegraphics[width=.49\textwidth]{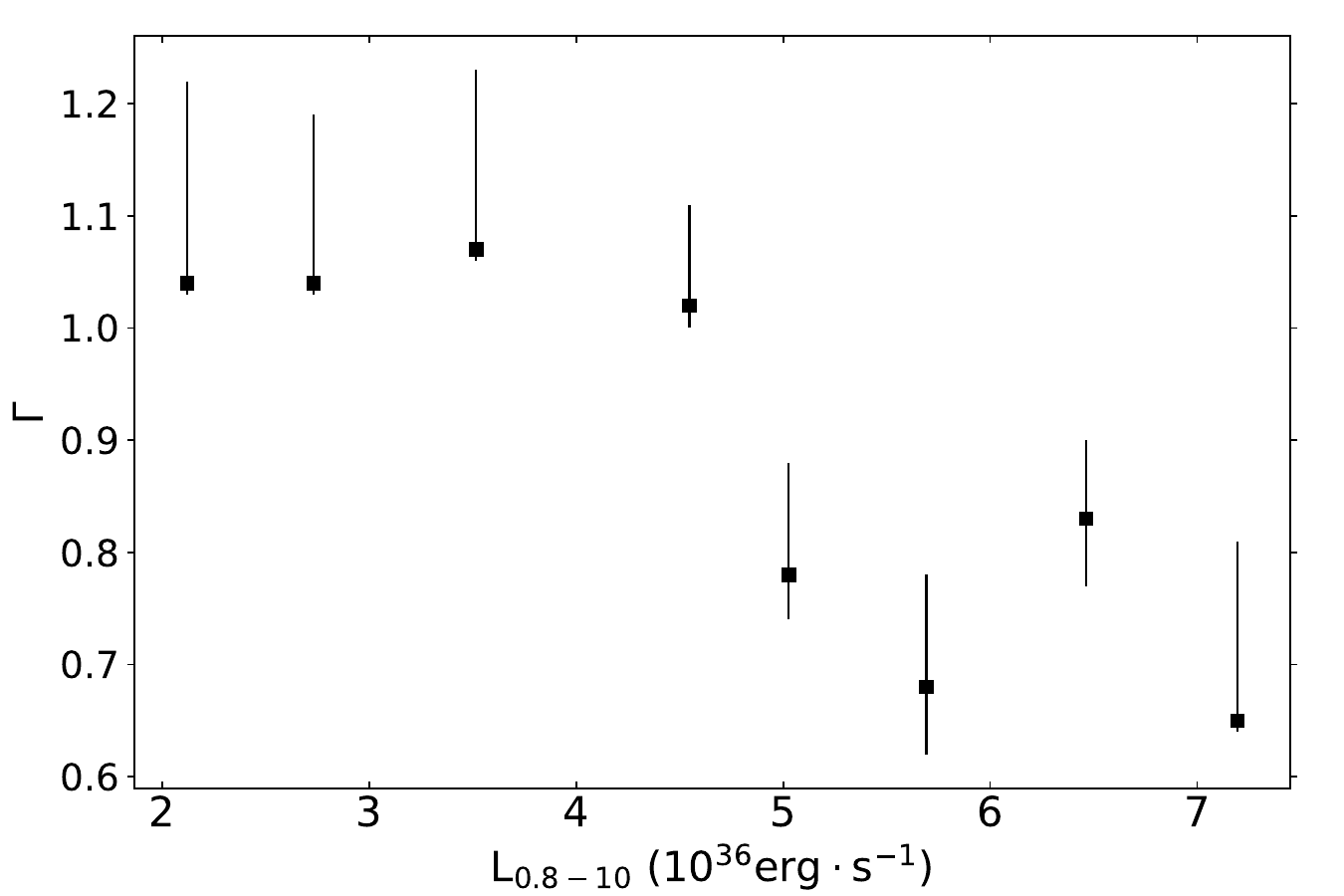}
    \caption{Photon index as a function of X-ray luminosity in the 0.8-10 keV range based on a distance of 8 kpc.}
    \label{fig:cyc_Flux}
\end{figure}

We presented a detailed spectral analysis of the accreting pulsar GRO~J1744--28 during its 2021 outburst, using observations by the \textrm{NICER} observatory. Thanks to NICER’s excellent sensitivity in the soft X-ray band, we confirm the presence of the previously reported cyclotron line at approximately 5 keV. Furthermore, we reveal a tentative evidence for a new cyclotron line at around 2 keV, detected in both the phase-averaged and phase-resolved spectra. The $\sim$2.4 ratio between the $\sim$5~keV and $\sim$2~keV features (if these are harmonic and fundamental CRSFs) significantly deviate from 2. This deviation could be interpreted by the theoretical prediction from the superposition model of cyclotron lines formed at different altitudes in the accretion column, as seen in Vela~X-1 \citep{2011ApJ...730..106N}. In the following discussion, we briefly analyse the variability of the spectral parameters derived from the phase-averaged spectra and summarize the implications of our findings.
 
From our phase-averaged spectral fits (see Table \cref{tab:ObsIDs_spec}), we find a hydrogen column density of $N_\mathrm{H} \approx 8.0 \times 10^{22}$~cm$^{-2}$. The photon index $\Gamma$ lies between 0.8 and 0.9, while the cut-off energy $E_\mathrm{cut}$ and folding energy $E_\mathrm{fold}$ lie in the $\sim$16--18~keV and $\sim$3--4~keV ranges, respectively. These continuum parameters differ slightly from previous fitting results \citep{2015MNRAS.449.4288D,2015MNRAS.452.2490D,2020A&A...643A.128K} (specifically, a larger $E_\mathrm{fold}$ and a smaller $E_\mathrm{cut}$). Nevertheless, these differences are expected, given the different models and energy bands used in the analyses. In the phase-averaged spectra, we detect the cyclotron line at $\sim$5~keV, although with a relatively weak depth. Additionally, an absorption feature at $\sim$2~keV is observed in both observations, which we interpret as a new CRSF if this line feature is real. Adopting the standard CRSF relation $E_\mathrm{cyc} \approx 11.57\,B_{12}$~keV (neglecting gravitational redshift $z$), the centroid energy of $\sim$2.0~keV corresponds to a surface magnetic field of $B \sim 1.8 \times 10^{11}$~G. This value is comparable to the field estimates obtained from spin-up measurements ($\sim 9 \times 10^{10}$~G) and the propeller-effect flux threshold ($\sim 2 \times 10^{11}$~G) reported in earlier studies \citep{2015ApJ...804...43Y, 1997ApJ...482L.163C}. The discrepancy between the magnetic field derived from the accretion torque and the CRSF may imply a quadratic magnetic field component, as suggested by \citet{2017MNRAS.469....2S,2019A&A...626A.106M}.

It is well-established that in many X-ray pulsars, the centroid energy of the CRSF correlates with X-ray luminosity, a phenomenon extensively reviewed by \citealt{2019A&A...622A..61S}. A positive correlation at low to moderate luminosities was first clearly demonstrated in Her~X-1 \citep{2007A&A...465L..25S} and has since been observed in several other systems, including Vela~X-1 \citep{2014ApJ...780..133F,2016MNRAS.463..185L,2022MNRAS.514.2805L}, Cep~X-4 \citep{2017A&A...601A.126V}, and V~0332+53 \citep{2017MNRAS.466.2143D,2018A&A...610A..88V}. Conversely, a negative correlation emerges above a critical luminosity, as first identified in V~0332+53 \citep{1990ApJ...365L..59M,2006MNRAS.371...19T} and later confirmed in 1A~0535+262. A clear anti-correlation thus appears above this threshold, and a positive correlation appears below it \citep{2015ApJ...806..193S,2021ApJ...917L..38K}.

This shift in correlation reflects a change between sub- and super-critical accretion regimes \citep{2022arXiv220414185M}. Below a critical luminosity, $L_{\mathrm{crit}}$ (the sub-critical regime), Coulomb interactions dominate the deceleration of the infalling material, and the height of the line-forming region decreases with increasing luminosity. Above $L_{\mathrm{crit}}$ (the super-critical regime), radiation pressure becomes dominant; a radiation-dominated shock decelerates the flow, forming an accretion column whose height increases with luminosity \citep{1976MNRAS.175..395B,2012A&A...544A.123B,2015MNRAS.447.1847M}. Since the local magnetic field strength varies with height in the column, the CRSF energy is therefore expected to increase with luminosity in the sub-critical regime and decrease in the super-critical regime. The positive correlation at low luminosities may also be influenced by Doppler effects \citep{2015MNRAS.454.2714M}.

In this work, the two NICER observations do not provide sufficient data to investigate such a luminosity-dependent CRSF energy in GRO~J1744--28. Nevertheless, the critical luminosity can be inferred theoretically. Following model of \citet{2012A&A...544A.123B}, the critical luminosity can be estimated for a dipole magnetic field as
\begin{equation}
    L_{\mathrm{crit}} \simeq 1.49 \times 10^{37} \ \mathrm{erg \ s^{-1}} \times B_{12}^{16/15},
\end{equation}
assuming canonical neutron-star parameters $M_{\mathrm{NS}} = 1.4\ M_{\odot}$ and $R_{\mathrm{NS}} = 10$~km. Using the magnetic field $B \sim 1.8 \times 10^{11}$~G ($B_{12} = 0.17$) derived from the CRSF, this gives $L_{\mathrm{crit}} \sim 2.4 \times 10^{36}$~erg~s$^{-1}$. This value is well below the observed luminosity of $(3-6)\times 10^{36}$~erg~s$^{-1}$ for a source distance of 8~kpc \citep{1997ApJ...486.1013A,1999ApJ...517..436N}, suggesting the source was in the super-critical regime. However, another theoretical calculation by \citet{2015MNRAS.447.1847M} yields $L_{\mathrm{crit}} \sim 10^{37}$ erg s$^{-1}$ for pure X-mode polarization and a column-radius scaling $l_0/l = 1.0$. This value is somewhat higher than other estimates, resulting in a sub-critical regime.

To further examine the variability of the continuum parameter $\Gamma$, we adopted the following approach. Given that the merged spectra provide only two data points and the cyclotron line parameters are poorly constrained in most individual observations, we fixed the line energy at 2 keV (as determined from the combined observation) and ignored the relatively weak feature at 5 keV. We then fitted each observation using the best-fit model described previously, thereby obtaining a robust measurement of the photon index. As depicted in Fig. \cref{fig:cyc_Flux}, the photon index $\Gamma$ --- varying between 0.6 and 1.1---exhibits a negative correlation with luminosity (Pearson coefficient r=-0.86, p=0.0066), indicating spectral hardening at higher flux levels. \citet{2015MNRAS.452.1601P} investigated the spectral hardness evolution in several transient X-ray pulsars and found that while hardness increases with flux at low luminosities, it saturates or even decreases above a critical luminosity. In contrast, our data show no evidence of saturation; the hardness ratio continues to increase across the observed luminosity range. This persistent hardening of $\Gamma$ suggests that GRO~J1744--28 was accreting in the sub-critical regime during these observations, with a theoretical critical luminosity likely exceeding $\sim (6-7) \times 10^{36}$~erg~s$^{-1}$.

\begin{acknowledgements}
We are grateful to the referee for the critical suggestions, which have helped improve the manuscript. This work is supported by the NSFC (12133007, 12173103 and 12261141691), the National Key Research and Development Program of China (Grant No. 2021YFA0718503), the Fundamental Research Funds for the Central Universities (010-63263126), and Science Research Project of Hebei Education Department (BJ2026091). LD acknowledges funding from German Research Foundation (DFG), Projektnummer 549824807. ST and JP acknowledge support by the Research Council of Finland, the Centre of Excellence in Neutron-Star Physics (project 374064). This work has made use of archival data provided by the High Energy Astrophysics Science Archive Research Center (HEASARC).
\end{acknowledgements}
%%%%%%%%%%%%%%%%%%%%%%%%%%%%%%%%%%%%%%%%%%%%%%%%%%

\bibliographystyle{aa}
\bibliography{refer}
%\label{lastpage}
\end{document}